\documentclass[12pt]{article}
\usepackage{epsfig}
\usepackage{lineno}
\usepackage{rotating}
\usepackage{lscape}
\usepackage{amsmath}
\usepackage[usenames]{color}

\usepackage{color}

\begin{document}

\begin{center}

\vspace*{1.0cm}

{\Large \bf{First search for 2$\varepsilon$ and
$\varepsilon\beta^+$ processes in $^{168}$Yb}}

\vskip 1.0cm

{\bf P.~Belli$^{a}$, R.~Bernabei$^{a,b,}$\footnote{Corresponding
author. {\it E-mail address:} rita.bernabei@roma2.infn.it
(R.~Bernabei).}, R.S.~Boiko$^{c,d}$, F.~Cappella$^{e}$,
V.~Caracciolo$^{f}$, R.~Cerulli$^{a}$, F.A.~Danevich$^{c}$,
M.L.~di~Vacri$^{f,g}$, A.~Incicchitti$^{e,h}$,
B.N.~Kropivyansky$^{c}$, M.~Laubenstein$^{f}$, S.~Nisi$^{f}$,
D.V.~Poda$^{c,i}$, O.G.~Polischuk$^{c}$, V.I.~Tretyak$^{c}$}

\vskip 0.3cm

$^{a}${\it INFN sezione Roma ``Tor Vergata'', I-00133 Rome, Italy}

$^{b}${\it Dipartimento di Fisica, Universit$\grave{a}$ di Roma ``Tor Vergata'', I-00133 Rome, Italy}

$^{c}${\it Institute for Nuclear Research, 03028 Kyiv, Ukraine}

$^{d}${\it National University of Life and Environmental Sciences of Ukraine, 03041 Kyiv, Ukraine}

$^{e}${\it INFN sezione Roma, I-00185 Rome, Italy}

$^{f}${\it INFN, Laboratori Nazionali del Gran Sasso, I-67100 Assergi (AQ), Italy}

$^{g}${\it Pacific Northwest National Laboratory, 902 Battelle
Blvd, 99354, Richland WA, US}

$^{h}${\it Dipartimento di Fisica, Universit$\grave{a}$ di Roma
``La Sapienza'', I-00185 Rome, Italy}

$^{i}${\it CSNSM, Univ. Paris-Sud, CNRS/IN2P3, Universit\'e
Paris-Saclay, 91405 Orsay, France}

\end{center}

\vskip 0.5cm

\begin{abstract}
The double-electron capture and the electron capture with positron
emission in $^{168}$Yb have been investigated for the first time
at the STELLA facility of the Gran Sasso underground laboratory
(Italy) measuring 371 g of highly purified ytterbium oxide placed
on the end-cap of a 465 cm$^3$ ultra-low-background high purity
Germanium detector (HPGe). No gamma associated to double beta
processes in $^{168}$Yb have been observed after 2074 h of data
taking. This has allowed setting the half-life limits on the level
of $\lim T_{1/2}\sim$ $10^{14}-10^{18}$ yr at 90\% C.L.
Particularly, a lower half-life limit on a possible resonant
neutrinoless double-electron capture in $^{168}$Yb to the $(2)^-$
1403.7 keV excited state of $^{168}$Er is set as
$T_{1/2}\geq1.9\times 10^{18}$ yr at 90\% C.L. Half-life limits
$T_{1/2}^{2\nu(0\nu)}\geq 4.5(4.3)\times10^{16}$ yr were set on
the $2\nu(0\nu)2\beta^-$ decay of $^{176}$Yb to the $2^+$ 84.3 keV
first excited level of $^{176}$Hf.

\end{abstract}

\vskip 0.4cm


\section{INTRODUCTION}

Double beta ($2\beta$) decays are rare processes of spontaneous
nuclear disintegration which changes the charge of nuclei by two
units. Double-electron capture ($2\varepsilon$), electron capture
with positron emission ($\varepsilon\beta^+$) and double-positron
decay ($2\beta^+$) result in the charge decreasing, while
double-electron decay ($2\beta^-$) leads to charge increasing.
Such processes are energetically allowed for 69 isotopes
\cite{Tretyak:2002} among 286 naturally occurring primordial
nuclides. Only 22 and 6 nuclides from the full list of 34
potentially $2\varepsilon$-active isotopes can also undergo
$\varepsilon\beta^+$ and $2\beta^+$ decays, respectively. Being
second order processes in the weak interaction, double beta decays
are characterized by extremely long half-lives, $\sim$10$^{19}$ yr
in the most favorable cases.

The main modes of $2\beta$ processes are two-neutrino ($2\nu$)
double beta decay, in which two (anti)neutrinos are also appearing
in the final state, and neutrinoless ($0\nu$) $2\beta$ decay. The
$2\nu 2\beta$ decay is a Standard Model (SM) process, already
observed in several nuclei with half-lives
$T^{2\nu2\beta}_{1/2}\sim 10^{19}-10^{24}$ yr
\cite{Barabash:2015,Tanabashi:2018}. The $0\nu2\beta$ decay requires
lepton number violation and a
Majorana nature of neutrino, therefore, it is a unique tool
to probe physics beyond the SM
(for more details see recent reviews
\cite{Vergados:2016,Delloro:2016,Bilenky:2015} and references
therein). The current most stringent half-life limits on $0\nu
2\beta^-$ decay are at the level of $\lim
T^{0\nu2\beta}_{1/2}\sim$ 10$^{24}$--10$^{26}$ yr
\cite{Tanabashi:2018}.

It should be emphasized that the above mentioned results are
obtained in $2\beta^-$ experiments, while the achievements in
investigations of $2\varepsilon$/$\varepsilon\beta^+$/$2\beta^+$
decays are much more modest. Indeed, there are only claims on the
detection of the double-electron capture in $^{130}$Ba (there is a
strong disagreement among the results of geochemical measurements,
summarized in \cite{Barabash:2015}) and $^{78}$Kr
\cite{Ratkevich:2017}. Moreover, the best experimental sensitivity
to the nuclear charge decreasing $2\beta$ processes is at the
level of $\lim
T^{2\varepsilon/\varepsilon\beta^+/2\beta^+}_{1/2}\sim$
10$^{21}$--10$^{22}$ yr (we refer reader to references 53--62 in
Ref. \cite{Belli:2018}). The reasons for this are: (i) small phase
space factors, i.e., suppressed decay probabilities (e.g., compare
the calculations for $2\beta^-$ \cite{Kotila:2012} with the ones
for $2\varepsilon$/$\varepsilon\beta^+$/$2\beta^+$ processes
\cite{Kotila:2013}); (ii) mostly very low natural abundances,
typically less than 1\%, with few exceptions \cite{Tretyak:2002};
(iii) experimental issues in the observation of a signature of the
most favorable process, i.e., $2\nu2\varepsilon$ capture
(difficulties in precisely and/or with high efficiency detection
of the emitted X-rays). At the same time, the searches for
$2\varepsilon$/$\varepsilon\beta^+$/$2\beta^+$ processes are
well-motivated by the following considerations: (i) the detection
of the $2\nu$ mode provides an important information for the
nuclear spectrometry and the measurement of the nuclear matrix
elements to be compared with theoretical calculations (see e.g.
\cite{Barabash:2015}); (ii) the observation of the $0\nu$ mode (or
a limit compatible to ones for the $0\nu2\beta^-$ decay) would
allow us to scrutinize the possible contribution of the
hypothetical right-handed currents to the mechanism of
neutrinoless double beta decay
\cite{Hirsch:1994,Deppisch:2012,Pas:2015}; (iii) the rate of
neutrinoless double-electron capture could be much faster thanks
to a resonant enhancement caused by a mass degeneracy between the
initial and final states
\cite{Winter:1955,Voloshin:1982,Bernabeu:1983,Sujkowski:2004}. All
together demand the development of experimental methods to
increase the current sensitivity to $2\beta$ processes. The main
goal of the present work is the investigation of double beta decay
processes in the ytterbium isotopes.

\nopagebreak
\begin{table}[!htb]
\caption{Potentially $2\beta$ decay active isotopes of ytterbium. }
\begin{center}
\begin{tabular}{|l|l|l|l|}
\hline
Double $\beta$              & Energy                            & Isotopic                          & Decay \\
transition                  & release (keV)                     & abundance (\%) \cite{Meija:2016}  & channel \\
 \hline
$^{168}$Yb$\to$$^{168}$Er   &   1409.27(25) \cite{Eliseev:2011} & 0.123(3)                          & $2\varepsilon$, $\varepsilon\beta^+$  \\
$^{176}$Yb$\to^{176}$Hf     &   1085.0(15)  \cite{Wang:2017}    & 12.995(83)                        & $2\beta^{-}$ \\
\hline
\end{tabular}
  \label{tab:YbDBDisotopes}
\end{center}
\end{table}

\begin{figure}[htbp]
\begin{center}
 \mbox{\epsfig{figure=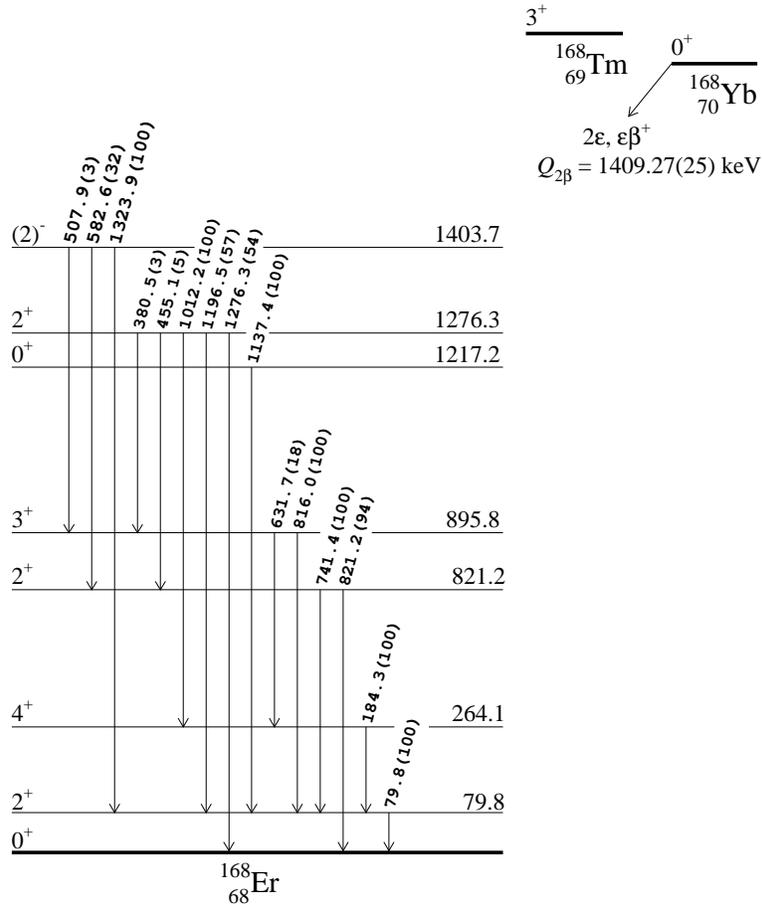,height=12.0cm}}
\vspace{-0.3cm} \caption{The simplified decay scheme of $^{168}$Yb
\cite{Baglin:2010}. The excited levels with spin 4 and higher
are omitted, except the 264.1 keV level. The energies of the levels
and of the emitted $\gamma$ quanta are in keV. Only $\gamma$
transitions with relative intensities of more than 2\% are
shown. The relative intensities of $\gamma$ quanta are rounded to
percent and given in parentheses.}
\end{center}
\label{fig:168Yb}
\end{figure}

\begin{figure}[htbp]
\begin{center}
 \mbox{\epsfig{figure=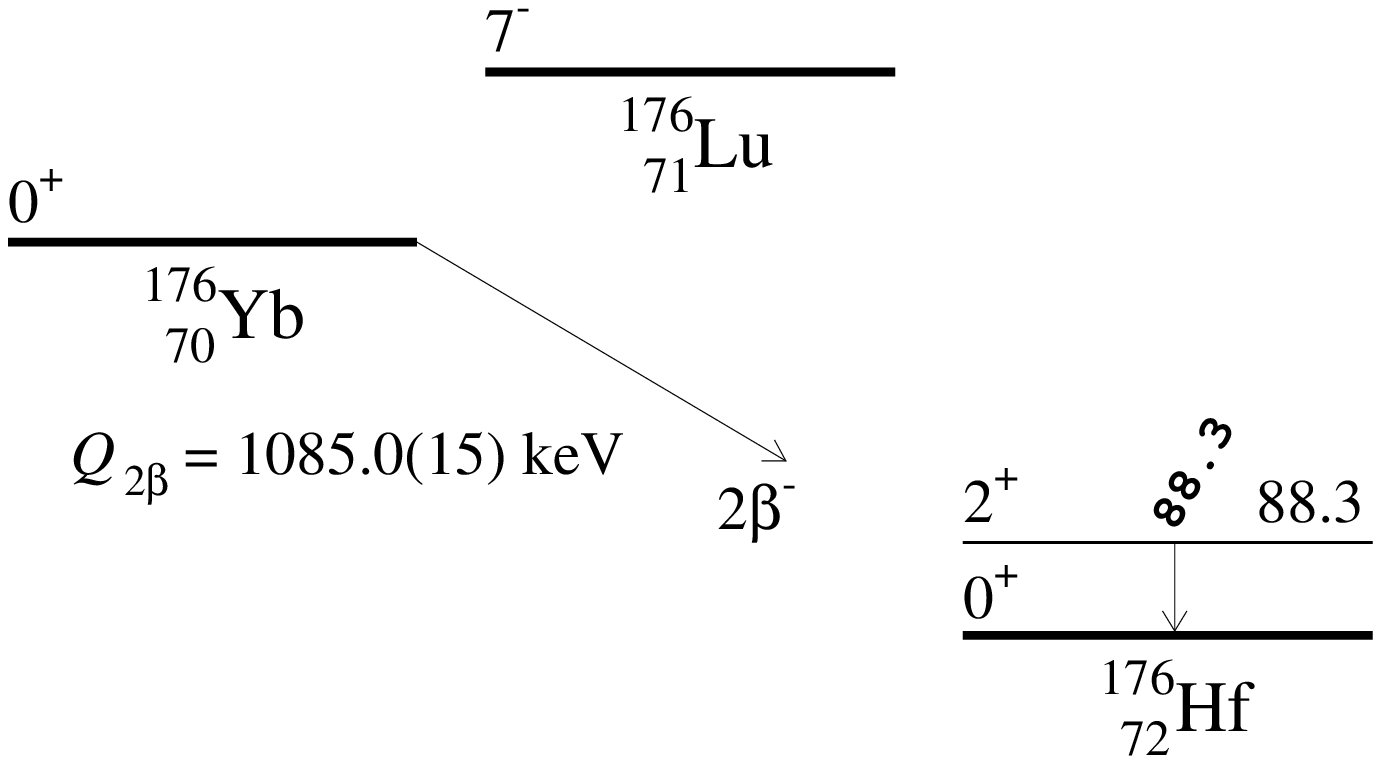,height=4.cm}}
\vspace{-0.3cm} \caption{The simplified decay scheme of $^{176}$Yb
\cite{Basunia:2006}. The energies of the excited level and of the
emitted $\gamma$ quantum are given in keV.} \label{fig:176Yb}
\end{center}
\end{figure}

Ytterbium contains two potentially double beta active nuclides:
$^{168}$Yb and $^{176}$Yb (see Table \ref{tab:YbDBDisotopes}).
Thanks to a reasonably high energy release ($Q_{2\beta}$), the
$2\varepsilon$ capture and $\varepsilon \beta^+$ decay to the ground
and to the several excited states of the daughter nucleus are energetically
allowed in $^{168}$Yb. A simplified scheme of the $^{168}$Yb
double beta decay transitions is presented in Fig.
\ref{fig:168Yb}. The theoretical predictions on the $^{168}$Yb
$2\nu 2\varepsilon$ capture rate \cite{Ceron:1999} suggest that
the detection of the decay to the ground state could be
feasible with an expected  half-life of 2$\times$10$^{23}$ yr,
while the transition to
the first 0$^+$ excited state (1217.2 keV) is experimentally
unreachable due to the extremely long expected half-life value
 of about 10$^{33}$ yr.  Moreover, the isotope $^{168}$Yb was also
proposed as a viable candidate for searches of resonant
$0\nu2\varepsilon$ transitions
\cite{Bernabeu:1983,Krivoruchenko:2011} and the expected
half-lives are in the range of 10$^{23}$--10$^{32}$ yr (assuming
the value of the effective Majorana neutrino mass equal to 1 eV)
\cite{Krivoruchenko:2011}. The recent high precision measurements
of the $^{168}$Yb $Q_{2\beta}$ value by the Penning-trap
mass-ratio \cite{Eliseev:2011} restrict possible resonant
transitions only to the 1403.7 keV excited level of $^{168}$Er. Up to
now, no experimental results on $^{168}$Yb double beta decay
half-lives were reported.

Another isotope of ytterbium, $^{176}$Yb, could reach via
$2\beta^-$ decay the ground and the first $2^+$ excited
level of $^{176}$Hf with energy 88.3 keV (see Fig.
\ref{fig:176Yb}). Until now, only the $2\beta^-$ transition to the 88.3 keV level
was investigated and the achieved lower half-life limit was
$1.6\times 10^{17}$ yr at 68\% C.L. \cite{Derbin:1996}. The
isotope $^{176}$Yb is also promising for the neutrino detection
\cite{Raghavan:1997} and the usage of a large-scale. Yb-containing
detector for such purpose would open possibilities for high-sensitivity
searches for $2\beta$ processes in $^{168}$Yb and $^{176}$Yb
\cite{Zuber:2000}. It is also worth noting that an
enrichment either in $^{168}$Yb or $^{176}$Yb to more than 15\%
and 97\%, respectively (a factor of 120 and 7.5, respectively)
was developed using a
laser isotope separation technology \cite{Park:2008}.

The present work describes the searches for double beta decay in a
sample of ytterbium oxide (Yb$_2$O$_3$) using the
ultra low background (ULB) gamma-ray spectrometry; therefore, only the
$2\beta$ processes resulting in the emission of X- and/or
$\gamma$-rays could be detected with the used experimental
technique. In addition, the work is also devoted to the study of
the possibilities of ytterbium purification from radioactive
elements, which is of particular interest for R\&D of
Yb-containing detectors of solar neutrinos and/or double beta
decay.

\section{Experiment}

\subsection{Purification of ytterbium oxide sample}
\label{sec:purification}

A sample of Yb$_2$O$_3$ powder ($>99.5\%$ TREO and $>99.999\%$ of
Yb$_2$O$_3$/TREO) has been purchased from Stanford Advanced
Materials Corporation. The material contamination has been
investigated by Inductively Coupled Plasma-Mass Spectrometry
(ICP-MS, model Element II from Thermo Fisher Scientific, Waltham,
Massachusetts, USA). The contaminants have been measured in Low
Resolution (LR) mode, while Si, Cl, Ca, Cr and Fe were measured in
Medium Resolution (MR) mode, and K was measured in High Resolution
(HR) settings, respectively, to overcome isobaric interference
issues. The results of the measurements are presented in Table
\ref{tab:YbICPMS}, where the data of the material supplier are
given too.

\begin{table}[!htb]
\caption{Concentration of impurities in the ytterbium oxide sample
measured by ICP-MS before and after the purification. Uncertainties on
the reported values are at the level of 30\% (Semi-Quantitative
mode of the ICP-MS measurements). Results of the ULB
gamma-spectrometry measurements (see text and Table
\ref{tab:YbHPGe}) are given in parentheses.}
\begin{center}
\begin{tabular}{|l|l|l|l|l|}
 \hline
 Element    & \multicolumn{4}{c|}{Concentration (ppm)} \\
 \cline{2-5}
 ~          & \multicolumn{2}{c|}{Before purification}  & Sediment after & After   \\
 \cline{2-3}
 ~          & Data of           & Present               & precipitation &  purification \\
 ~          & supplier          & measurement           &               &  \\
 \hline
 Si         &  10               & $<20$                 & -             & - \\
 \hline
 Cl         &  50               & -                     & -             & - \\
 \hline
 K          &  -                & $<2$ (0.8)            & 0.3           & 0.3 (0.3) \\
 \hline
 Ca         & 5                 & $<80$                 & -             & - \\
 \hline
 Cr         & -                 & $<0.2$                & -             & - \\
 \hline
 Fe         & 1                 & 1.4                   & 1.8           & 0.09 \\
 \hline
 Co         &  -                & 0.058                 & -             & - \\
 \hline
 Cu         & $<5$              & 0.6                   & -             & - \\
 \hline
 Cs         &  -                & $<0.01$               & -             & - \\
 \hline
 La         &  $<1$             & 0.01                  & 0.002         & 0.006 ($<0.15$)\\
 \hline
 Ce         &  $<1$             & 0.05                  & -             & - \\
 \hline
 Eu         &  $<1$             & $<1$                  & -             & - \\
 \hline
 Lu         &  1                & -   (7.6)             & $<15$         & $<15$ (7.8) \\
 \hline
 Pb         &  -                & 0.7                   & 0.3           & 0.5 \\
 \hline
 Th         &  -                & 0.013 (0.019)         & 0.2           & $<0.0005$ (0.0003) \\
 \hline
 U          &  -                & 0.0044 ($<0.012$)     & 0.0014        & $0.001$ ($<0.001$) \\
 \hline
\end{tabular}
\end{center}
\label{tab:YbICPMS}
\end{table}

The radioactive contamination of the material has been studied
using high-purity germanium (HPGe) $\gamma$-ray spectrometry at
the STELLA (SubTErranean Low Level Assay) facility of the Gran
Sasso underground laboratory (Italy). A 340 h long measurement of
a 627 g Yb$_2$O$_3$ sample has been performed using a p-type
detector, GePaolo (active volume of 518 cm$^3$ and relative
efficiency, with respect to a $3'' \times 3''$ sodium iodine
detector, of 113\% \cite{Laubenstein:2017}). The detection
efficiencies to $\gamma$ quanta in the full energy peak were
simulated using the GEANT4 package
\cite{Agostinelli:2003,Allison:2006}, and especially for the
double beta decay processes  by using the event generator DECAY0
\cite{Ponkratenko:2000,DECAY0}. The intensities of the $\gamma$
peaks in the energy spectra accumulated with and without the
Yb$_2$O$_3$ powder have been compared to estimate the activities
(or upper limits on the activities) of the radioactive impurities.
The activity of the $^{228}$Ra was estimated by analysis of the
$\gamma$ quanta emitted by its daughter $^{228}$Ac, the activity
of $^{228}$Th was derived from $^{212}$Pb, $^{212}$Bi, and
$^{208}$Tl, the limit on $^{226}$Ra was obtained from analysis of
$^{214}$Pb and $^{214}$Bi, the activities of $^{235}$U, $^{238}$U,
$^{234}$Th and $^{234m}$Pa were found by analysis of the $\gamma$
quanta expected in the decays of the nuclides. The HPGe
measurements of the initial material (denoted in Tables
\ref{tab:YbICPMS} and \ref{tab:YbHPGe} as ``Before purification'')
show traces of potassium ($^{40}$K), lutetium ($^{176}$Lu), radium
($^{228}$Ra) and thorium ($^{228}$Th). It should be stressed that
the activities of $^{40}$K,  $^{176}$Lu and $^{228}$Th do
contradict neither the data of the material supplier nor ICP-MS
analysis of potassium, thorium and lutetium.

 \nopagebreak
\begin{table}[!htb]
 \caption{
Radioactive contamination of the ytterbium oxide samples measured
with ultra-low background HPGe $\gamma$ detector with active
volume of 518 cm$^3$ (627 g sample, before the purification) and
in the course of the experiment with the help of HPGe $\gamma$
detector with active volume of 465 cm$^3$ (371 g, after the
purification). The uncertainties are given with $\approx 68\%$
C.L., while the upper limits were evaluated with 90\% C.L.
 }
\begin{center}
\begin{tabular}{|l|l|l|l|}
 \hline
  Chain     & Nuclide       & \multicolumn{2}{c|}{Activity (mBq/kg)} \\
 \cline{3-4}
  ~         & ~             & Before purification   & After purification \\
\hline
 ~          & $^{40}$K      & $26\pm9$                      & $9\pm4$ \\
 ~          & $^{137}$Cs    & $\leq1.8$                     & $1.1\pm0.2$ \\
 ~          & $^{138}$La    & -                             & $\leq0.12$ \\
 ~          & $^{152}$Eu    & -                             & $\leq0.59$ \\
 ~          & $^{154}$Eu    & -                             & $\leq0.24$ \\
 ~          & $^{176}$Lu    & $410\pm30$                    & $420\pm30$ \\
 \hline
 $^{232}$Th & $^{228}$Ra    & $88\pm5$                      & $2.6\pm0.7$ \\
 ~          & $^{228}$Th    & $75\pm4$                      & $1.1\pm0.5$ \\
 \hline
 $^{235}$U  & $^{235}$U     & $\leq13$                      & $\leq2.5$ \\
 ~          & $^{227}$Th    & -                             & $\leq4.1$ \\
 ~          & $^{223}$Ra    & -                             & $\leq4.6$ \\
 ~          & $^{211}$Pb    & -                             & $\leq3.7$ \\
 ~          & $^{207}$Tl    & -                             & $\leq34$ \\
 \hline
 $^{238}$U  & $^{234}$Th    & $\leq2100$                    & $\leq240$ \\
 ~          & $^{234m}$Pa   & $\leq150$                     & $\leq13$ \\
 ~          & $^{226}$Ra    & $\leq2.8$                     & $\leq0.83$ \\
\hline
\end{tabular}
\end{center}
\label{tab:YbHPGe}
\end{table}

To reduce the observed radioactive contamination,
purification of the material was performed using a combination
of the sedimentation (precipitation) and liquid-liquid extraction
\cite{Boiko:2017} methods.

As a first step the ytterbium oxide sample was dissolved in a
solution of $\leq 67\%$ nitric acid (``HyperPur'', Alfa Aesar) in
deionized water (18.2 M$\Omega \cdot$ cm):

\begin{equation}
\mbox{Yb}_2\mbox{O}_3 + \mbox{6HNO}_3 = \mbox{2Yb(NO}_{3})_{3} +
 3\mbox{H}_2\mbox{O}.
 \label{eg:1}
\end{equation}

\noindent The analysis of the obtained solution has shown
concentrations of 1.59 M for Yb(NO$_3$)$_3$ and of 0.9 M for the
residual HNO$_3$.

Then, the co-precipitation technique using fractional
sedimentation of carrying ytterbium hydroxide Yb(OH)$_3$ was
applied to remove low-soluble impurities from the solution, such as
thorium and iron\footnote{Purification of ytterbium oxide from
iron (and other transition metals) was motivated by prospects of a future
Yb-containing crystal scintillators development.} hydroxides.
Gaseous ammonia, produced from ammonium hydroxide solution
($20.0\%-30.0\%$ NH$_3$ basis, ACS reagent, Sigma-Aldrich), was
injected into the acidic solution (till $\mbox{pH}=6.5$) to
obtain ytterbium hydroxide which partially precipitated
at this
condition absorbing Th and Fe:

\begin{equation}
 \mbox{Yb(NO}_3)_3+\mbox{3NH}_3+\mbox{3H}_2\mbox{O}=\mbox{Yb(OH)}_3\downarrow +~3\mbox{NH}_4\mbox{NO}_3,
 \label{eg:2}
\end{equation}

\begin{equation}
 \mbox{Th}^{4+}+\mbox{Fe}^{3+}+7\mbox{NH}_3+7\mbox{H}_2\mbox{O}\rightarrow \mbox{Th(OH)}_4\downarrow
 +\mbox{Fe(OH)}_3\downarrow + \mbox{7NH}_{4}^{+}.
 \label{eg:3}
\end{equation}

The obtained sediment (135 g, that is 21.6\% of the initial amount
of the material) was converted to Yb$_2$O$_3$ form and analyzed by
ICP-MS. The results of the measurement are presented in Table
\ref{tab:YbICPMS}. The data show accumulation of Fe and
especially of Th in the sediment, and therefore,  confirm the ability of the
sedimentation method to reduce contamination of the impurities in
the initial Yb$_2$O$_3$.

The remaining aqueous solution of ytterbium nitrate was acidified to
$\mbox{pH}\sim1$ and purified by the liquid-liquid extraction
technique. The liquid-liquid extraction method has been proven to
be the most effective technique for the purification of
lanthanides from uranium and thorium traces
\cite{Boiko:2017,Belli:2014,Belli:2018}. A solution of 0.1 M
tri-n-octylphosphine oxide (TOPO, 99\%, Acros Organics) diluted in
toluene ($\leq 99.7\%$, ACS reagent, Sigma-Aldrich) has been
used as non-polar organic phase for the extraction of U and Th
from the polar aqueous solution.

The two immiscible liquids were put into a separation funnel in
volumetric ratio 1:1 and were shaken for several minutes. Uranium and
thorium moved to the organic phase forming organometallic
complexes by interaction with TOPO:

\begin{equation}
 \mbox{Yb(Th,U)(NO}_3)_{3(aq.)}+\mbox{nTOPO}_{(org.)}=\mbox{Yb(NO}_3)_{3(aq.)}
+ \mbox{[(Th,U)}\bullet \mbox{nTOPO]}_{(org.)}.
 \label{eg:4}
\end{equation}

The purified aqueous solution was separated from the organic phase,
and then ytterbium was precipitated from the solution in form of
hydroxide by using ammonia gas. We expect that at this stage of
the purification alkali and alkali-earth cationic impurities
partially remained in the supernatant liquid. Finally, the
obtained sediment was separated, dried and annealed at $\approx
900^{\circ}\mbox{C}$ over several hours. As a result, 420 g of
purified material have been obtained that is 67\% of the initial
material.

\subsection{Measurements}

\begin{figure}[htb]
\begin{center}
 \mbox{\epsfig{figure=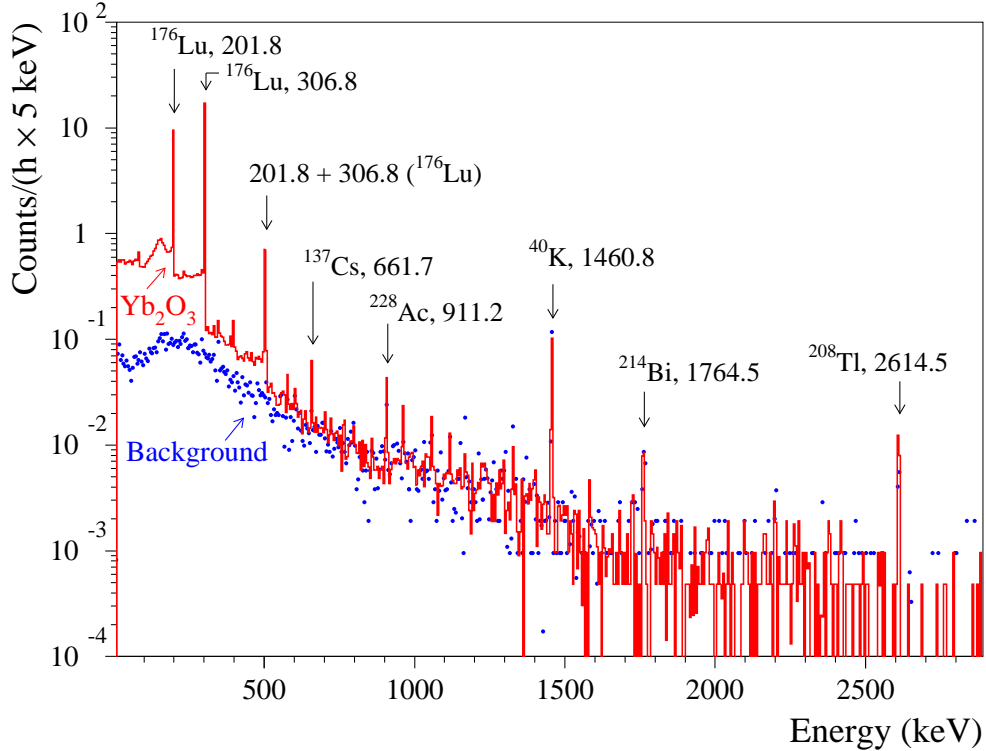,height=10.0cm}}
\vspace{-0.3cm} \caption{The energy spectra measured with the
ULB HPGe $\gamma$ spectrometer with the
Yb$_2$O$_3$ sample over 2074 h (Yb$_2$O$_3$) and without the sample
over 1046 h (Background). The energies of the most prominent
$\gamma$ peaks are given in keV.}
 \label{fig:bg}
\end{center}
\end{figure}

A 371 g sample of the purified Yb$_2$O$_3$ was
placed in a cylindric box of polystyrene (diameter 90 mm, height 50 mm)
 and placed on the end cap of the HPGe detector
GeCris (active volume 465 cm$^3$, 120 \% relative efficiency  with respect to a 3" $\times$ 3" sodium iodine detector)
\cite{Laubenstein:2017}) of the STELLA facility
of the Gran Sasso underground laboratory. The $\gamma$
spectrometer is shielded all around by $\approx$25 cm
of low radioactive lead, $\approx$5 cm copper, and $\approx$2.5 cm
Roman lead in the innermost part. The set-up is enclosed in an
air-tight poly(methyl methacrylate) box and flushed with high
purity nitrogen gas to reduce as much as possible the environmental radon
concentration. The energy dependence of the detector resolution is
described by the following function: FWHM(keV)~$=\sqrt{1.41+0.00197\times E_{\gamma}}$,
where $E_{\gamma}$ is in keV. The energy spectrum with the
Yb$_2$O$_3$ sample was gathered over 2074 h and the background
spectrum was measured over 1046 h. The spectra normalized on time of
measurements, are shown  in Fig. \ref{fig:bg}.

There was still some excess in the $\gamma$ peaks of $^{40}$K,
$^{137}$Cs, $^{176}$Lu, and $^{232}$Th daughters in the data
of the purified Yb$_2$O$_3$ sample. The estimations of the
radioactive contamination of the sample after purification are
presented in Table \ref{tab:YbHPGe}. Unfortunately, the
contamination in lutetium remained the same due to the high
chemical affinity of ytterbium and lutetium, while the
concentrations of potassium, radium and thorium was reduced by
factors of 3, 34 and 68, respectively.

\section{Searches for $2\beta$ processes in $^{168}$Yb and $^{176}$Yb}

The energy spectrum of the Yb$_2$O$_3$ sample (see
Fig. \ref{fig:bg}) does not contain peculiarities which could be
ascribed to double beta decay processes in ytterbium isotopes.
Therefore, the data have been analyzed estimating
lower half-life limits for the $2\varepsilon$ and the
$\varepsilon\beta^+$ decay of $^{168}$Yb and $2\beta^-$ decay of
$^{176}$Yb using the following formula:

\begin{center}
$\lim T_{1/2} = \ln 2 \cdot N \cdot  t \cdot \eta / \lim S,$
\end{center}

\noindent where $N$ is the number of potentially $2\beta$ active
nuclei in the Yb$_2$O$_3$ sample, $t$ is the time of measurement,
$\eta$ is the detection efficiency for a considered $2\beta$
channel, and $\lim S$ is the upper limit on the number of events
of the effect searched for, which can be excluded at a given
confidence level (C.L.).

The EGSnrc-based Monte Carlo simulations \cite{Kawrakow:2003} with
initial kinematics given by the DECAY0 event generator
\cite{Ponkratenko:2000,DECAY0} have been performed to calculate
the detection efficiencies. The Monte Carlo simulations have been
shown to be accurate by comparison with experimental data
obtained with a $^{133}$Ba $\gamma$ calibration source. The areas of the
$^{133}$Ba $\gamma$ peaks with energies 276 keV, 303 keV, 356 keV
and 384 keV agree with the experimental data within 1\%, while the
low energy peak at 81 keV is smaller than the experimental one by
28\%. The discrepancy for the low energy $\gamma$ quanta can be
explained by difficulties to reconstruct precisely the
experimental geometry (in particular, the effect could be due to
the source holder that caused some additional absorption of low
energy $\gamma$ quanta).

By fitting the data with the models accounting for the effect
searched for and for the background, the $\lim S$ has been
estimated according to the Feldman-Cousins recommendations
\cite{Feldman:1998}. All the $\lim S$, and thus the half-life
limits in the present work, are given at 90\% C.L. The
contributions of systematic errors, e.g. uncertainties of the
energy calibration and resolution, the number of $2\beta$ active
nuclei, and time of measurements are negligible in comparison to
the statistical fluctuations of the excluded peak areas.
Uncertainties of the Monte Carlo calculated detection efficiencies
may have some merit at the energies less than $\sim270$ keV as it
was demonstrated by the comparison of the Monte Carlo simulation
and experimental data obtained with the $^{133}$Ba source.
Nevertheless, we decided to include only statistical errors coming
from the data fluctuations in the half-life limits estimations.

\subsection{Double-electron capture in $^{168}$Yb}

\begin{figure}[htb]
 \begin{center}
  \mbox{\epsfig{figure=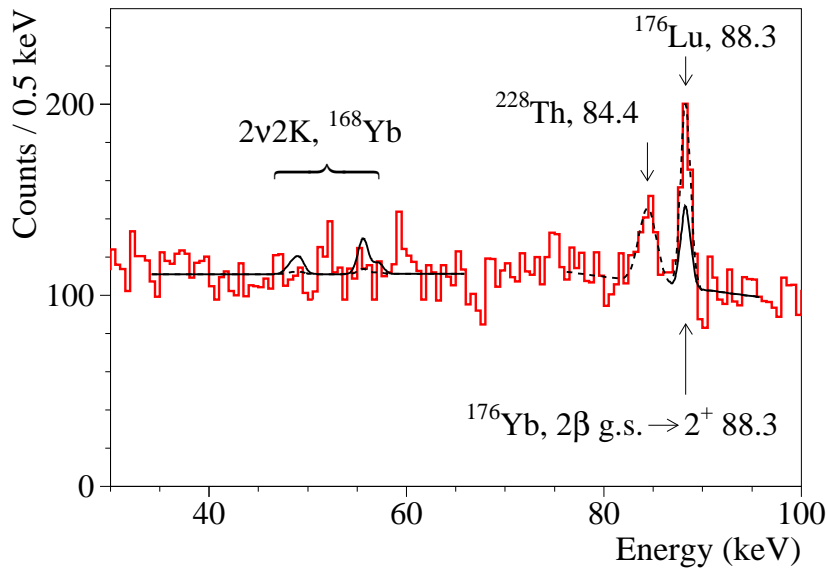,height=7.5cm}}
 \vspace{-0.3cm}
\caption{A low energy part of the spectrum accumulated by the
ultra-low background $\gamma$ spectrometer with the Yb$_2$O$_3$
sample over 2074 h. The approximation function and the excluded
effect of $2\nu2K$ capture in $^{168}$Yb are shown by dashed and
solid lines, respectively. The fit of the data in the vicinity of
the $2\beta^-$ transition of $^{176}$Yb to the first excited level
(88.3 keV) of $^{176}$Hf (dashed line) and the excluded peak of
the effect (solid line) are also shown. The energies of the
$\gamma$ peaks are given in keV.}
 \label{fig:2n2K}
 \end{center}
 \end{figure}

A cascade of X rays and Auger electrons with energies $47.7-57.5$
keV is the expected signature of the $2\nu2K$ capture in
$^{168}$Yb with the transition to the ground state of $^{168}$Er.
In the analysis we considered the most intense X rays of erbium
\cite{Firestone:1998}: $E_1=48.2$ keV (the intensity is
$I_1=27.0\%$), $E_2=49.1$ keV ($I_2=47.5\%$), $E_3=55.5$ keV
($I_3=5.1\%$), $E_4=55.7$ keV ($I_4=9.8\%$), and $E_5=57.1$ keV
($I_5=3.3\%$). The detection efficiency for $\gamma$ quanta in the
energy interval $49-57$ keV was calculated using the EGSnrc-based
Monte Carlo code. The detection efficiency at such a low energy
strongly depends on energy and can be approximated in the energy
interval $48-57$ keV as following: $\eta=7.639\times
10^{-3}-3.151\times 10^{-4}\times E+3.260\times 10^{-6}\times
E^2$, where $E$ is in keV. The whole distribution of the expected
X-ray peaks was built by accounting for the intensities $I_i$ and
the efficiencies at the energies $E_i$. Then, the energy spectrum
of the Yb$_2$O$_3$ sample, in the region of interest, was fitted
with the sum of the five Gaussians (normalized to the total area
equal 1) and a first-degree polynomial to model the continuous
background (see Fig. \ref{fig:2n2K}). The best fit
($\chi^2/$n.d.f.$=55.6/62=0.864$, where n.d.f. is number of
degrees of freedom), in the energy interval ($34-66$) keV, gives
($21.8 \pm 79.1$) counts for the $2\nu2K$ effect. According to
\cite{Feldman:1998}, the $\lim S$ value was calculated to be 152
counts. The detection efficiency for the five peaks was calculated
as $\Sigma I_i \times \eta_{i}=5.82\times10^{-5}$, where
$\eta_{i}$ is the detection efficiency at energy $E_i$. Finally,
considering that $1.39\times10^{21}$ nuclei of $^{168}$Yb are in
the Yb$_2$O$_3$ sample, one obtains the following half-life limit
for the $2\nu2K$ capture in $^{168}$Yb:
$T_{1/2}\geq8.7\times10^{13}$ yr. The simulated efficiency at the
81 keV for $^{133}$Ba source was lower than the measured one, thus
our $T_{1/2}$ limit for the $2\nu2K$ process is rather
conservative.

In case of $2\nu2\varepsilon$ decays of $^{168}$Yb to excited
states of $^{168}$Er, only the transitions to $0^+$ and $2^+$
levels were considered. The spectrum with the Yb$_2$O$_3$ sample
has been fitted in the various energy intervals, where the most
intense $\gamma$ quanta are expected to be emitted in the
de-excitation processes (see Fig. \ref{fig:168Yb}). The background
models were built from a Gaussian function describing the effect
(a peak with the energy of the considered de-excitation $\gamma$
quanta), a continuous background and additional Gaussian(s) if it
was also needed to approximate neighboring $\gamma$ peak(s) coming
from environmental radioactivity and/or internal contamination of
the Yb$_2$O$_3$. The derived half-life limits are reported in
Table \ref{table:DBDlimits}.


\begin{table*}[!htbp]
\vspace{-0.7cm} \caption{Half-life limits on 2$\beta$ processes in
$^{168}$Yb and $^{176}$Yb, established in the present work. All
the limits are given with 90\% C.L.} \footnotesize
\begin{center}
\resizebox{1.20\textwidth}{!}{
\begin{tabular}{|l|l|l|l|l|l|l|}
\hline
 Process                    & Decay     & Level of      & Signature     & Detection             & $\lim S$  & Half-life \\
 of 2$\beta$ decay          & mode      & daughter      & $E_\gamma$    & efficiency            & (counts)  & limit (yr) \\
 ~                          & ~         & nucleus       & (keV)         &                       &           &  \\
 ~                          & ~         & (keV)         &               &                       &           &  \\
 \hline
 $^{168}$Yb$~\to$$^{168}$Er & ~         & ~             & ~             & ~                     &  ~        &  ~\\
 $2K$                       & $2\nu$    & g.s.          & 48--57        & $5.818\times10^{-5}$  & 152       & $\geq8.7\times10^{13}$ \\
 $2\varepsilon$             & $2\nu$    & $2^{+}~79.8$  & 79.8          & $3.974\times10^{-4}$  & 40        & $\geq2.3\times10^{15}$ \\
 $2\varepsilon$             & $2\nu$    & $2^{+}~821.2$ & 821.2         & $1.275\times10^{-2}$  & 6.6       & $\geq4.4\times10^{17}$ \\
 $2\varepsilon$             & $2\nu$    & $0^{+}~1217.2$& 1137.4        & $2.311\times10^{-2}$  & 3.5       & $\geq1.5\times10^{18}$ \\
 $2\varepsilon$             & $2\nu$    & $2^{+}~1276.3$& 1276.3        & $5.296\times10^{-3}$  & 1.4       & $\geq8.6\times10^{17}$ \\
 ~                          & ~         & ~             & ~             & ~                     & ~         & ~                     \\
 $2K$                       & $0\nu$    & g.s.          & 1294.0--1294.5 & $2.180\times10^{-2}$ & 6.8       & $\geq7.3\times10^{17}$ \\
 $KL$                       & $0\nu$    & g.s.          & 1341.8--1343.7 & $2.142\times10^{-2}$ & 7.1       & $\geq6.9\times10^{17}$ \\
 $2L$                       & $0\nu$    & g.s.          & 1389.5--1392.8 & $2.101\times10^{-2}$ & 3.9       & $\geq1.2\times10^{18}$ \\
 $2\varepsilon$             & $0\nu$    & $2^{+}~79.8$  & 79.8          & $7.698\times10^{-5}$  & 40        & $\geq4.4\times10^{14}$ \\
 $2\varepsilon$             & $0\nu$    & $2^{+}~821.2$ & 821.2         & $1.126\times10^{-2}$  & 6.6       & $\geq3.9\times10^{17}$ \\
 $2\varepsilon$             & $0\nu$    & $0^{+}~1217.2$& 1137.4        & $2.310\times10^{-2}$  & 3.5       & $\geq1.5\times10^{18}$ \\
 $2\varepsilon$             & $0\nu$    & $2^{+}~1276.3$& 1276.30       & $5.307\times10^{-3}$  & 1.4       & $\geq8.6\times10^{17}$ \\
 Resonant $M_1M_1$          & $0\nu$    & $(2)^{-}~1403.7$& 1323.9      & $1.590\times10^{-2}$  & 1.9       & $\geq1.9\times10^{18}$ \\
 $\varepsilon\beta^+$       & $2\nu$    & g.s.          & 511           & $6.277\times10^{-2}$  & 51        & $\geq2.8\times10^{17}$ \\
 $\varepsilon\beta^+$       & $2\nu$    & $2^{+}~79.8$  & 511           & $6.274\times10^{-2}$  & 51        & $\geq2.8\times10^{17}$ \\
 $\varepsilon\beta^+$       & $0\nu$    & g.s.          & 511           & $6.190\times10^{-2}$  & 51        & $\geq2.8\times10^{17}$ \\
 $\varepsilon\beta^+$       & $0\nu$    & $2^{+}~79.8$  & 511           & $6.223\times10^{-2}$  & 51        & $\geq2.8\times10^{17}$ \\
 ~                          & ~         & ~             & ~             & ~                     & ~         & ~                     \\
  $^{176}$Yb$~\to^{176}$Hf  & ~         & ~             & ~             & ~                     & ~         & ~                     \\
 $2\beta^{-}$               &$2\nu$     & $2^+$ 88.3    & 88.3          & $1.958\times10^{-4}$  & 106       & $\geq4.5\times10^{16}$  \\
 $2\beta^{-}$               &$0\nu$     & $2^+$ 88.3    & 88.3          & $1.907\times10^{-4}$  & 106       & $\geq4.3\times10^{16}$  \\
 \hline
\end{tabular}
 }
\end{center}
\label{table:DBDlimits}
\end{table*}
\normalsize

\begin{figure}[htb]
 \begin{center}
  \mbox{\epsfig{figure=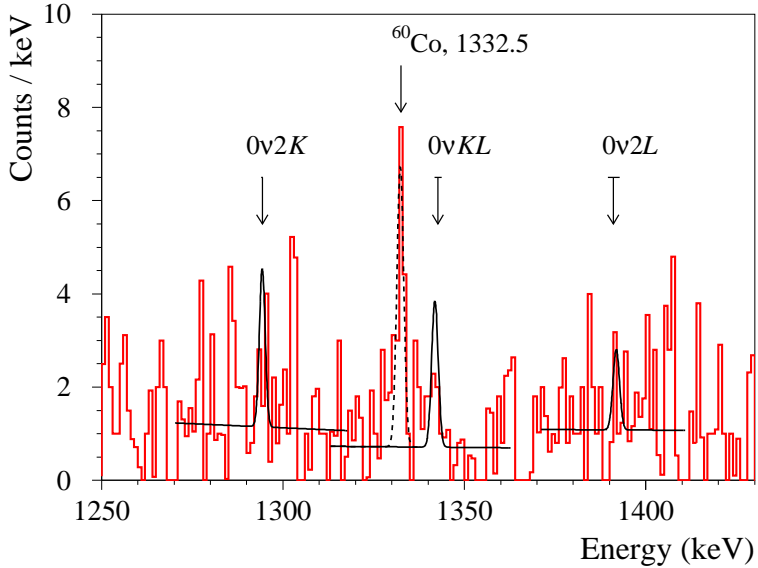,height=7.5cm}}
\vspace{-0.3cm} \caption{An interval of the energy spectrum
accumulated with the Yb$_2$O$_3$ sample over 2074 h, where the
$\gamma$ peaks from the $0\nu 2K$, $0\nu KL$, and $0\nu 2L$
captures in $^{168}$Yb to the ground state of $^{168}$Er are
expected. The 90\% C.L. excluded peaks are shown by solid lines.
The fit of the background peak of $^{60}$Co with energy 1332.5 keV
is shown by the dashed line.}
 \label{fig:0n2e}
 \end{center}
 \end{figure}

In the searches for neutrinoless double-electron capture in
$^{168}$Yb to the ground state of $^{168}$Er, we have considered only
captures from $K$ and/or $L$ shells (as the most probable). It was
also supposed that the energy excess in the process is taken away
by (bremsstrahlung) $\gamma$ quanta with an energy
$E_{\gamma}=Q_{2\beta}-E_{b1}-E_{b2}$, where $E_{b1}$ and $E_{b2}$
are the binding energies of the captured electrons on the atomic
shells of the daughter erbium atom. The energy spectrum measured
with the Yb$_2$O$_3$ sample was then fitted by a model constructed
in the same way as in the above-mentioned analysis of the
$2\nu2\varepsilon$ decays of $^{168}$Yb to the excited states of
$^{168}$Er. However, the energy of the Gaussian function used to
describe the effect was varied according to the uncertainty of the
$Q_{2\beta}$ value and, for the captures involving the $L$ shells,
to the difference between the binding energies of $L_1$ and $L_3$
shells. The fits to the energy spectrum find ($3.2 \pm 2.2$),
($3.2\pm2.4$), and ($1.6\pm1.4$) counts for the $0\nu 2K$, $0\nu
KL$, and $0\nu 2L$ peaks respectively and the corresponding $\lim
S$ values are 6.8, 7.1 and 3.9 counts. The excluded peaks of the
$0\nu$ double-electron captures in $^{168}$Yb to the ground state
of $^{168}$Er are shown in Fig. \ref{fig:0n2e}, while the obtained
lower half-life limits on these $2\beta$ processes are given in
Table \ref{table:DBDlimits}.

The $2\varepsilon$ decay of $^{168}$Yb is also allowed to excited
levels of $^{168}$Er, however the $0\nu$ and $2\nu$ modes cannot
be distinguished by $\gamma$-ray spectrometry. The
detection efficiencies for $0\nu$ and $2\nu$ modes are slightly
different due to emission of additional $\gamma$ quanta in the
$0\nu2\varepsilon$ decay with energy
$E_\gamma=Q_{2\beta}-E_{b1}-E_{b2}-E_{exc}$, where $E_{exc}$ is the energy
of the excited level of $^{168}$Er. So, using the results of fits
in the energy intervals where intense $\gamma$ peaks are expected,
we obtained limits on the $0\nu2\varepsilon$ decays of $^{168}$Yb
to the $0^{+}$ and $2^{+}$ excited levels of $^{168}$Er (see Table
\ref{table:DBDlimits}).

The rate of the $0\nu2\varepsilon$ capture in $^{168}$Yb to the
excited levels of $^{168}$Er could be resonantly enhanced thanks
to the possible degeneracy of the energy release. Taking into
account the recent precise measurements of the $^{168}$Yb
$Q_{2\beta}$ value \cite{Eliseev:2011} and the nuclear structure
of $^{168}$Er \cite{Baglin:2010}, the lowest degeneracy
$Q_{2\beta} - E_{b1} - E_{b2} - E_{exc}=1.16$ keV can be achieved
in the $0\nu M_1M_1$ decay of $^{168}$Yb to the $(2)^{-}$ 1403.7
keV level of $^{168}$Er. However, no substantial resonant
enhancement of the decay probability is foreseen for the
degeneracy on the 1-keV-level (e.g., see Fig.~10 in
\cite{Belli:2012}). Moreover, the capture of two electrons from
higher orbitals, as well as the transition between 0$^+$ and
(2)$^-$ states\footnote{Not precise knowledge of the $^{168}$Er
nuclear structure leaves a possibility of more favorable ($0^+
\rightarrow 0^-$) transition.}, additionally suppress the
probability of the decay. Nevertheless, we have performed the
search for this process too and the absence of 1323.9 keV
de-excitation $\gamma$ quanta in the background spectrum of the
Yb$_2$O$_3$ sample was used to set a half-life limit
$T_{1/2}\geq1.9\times10^{18}$ yr.

\subsection{Electron capture with positron emission in $^{168}$Yb}

\begin{figure}[htb]
\begin{center}
 \mbox{\epsfig{figure=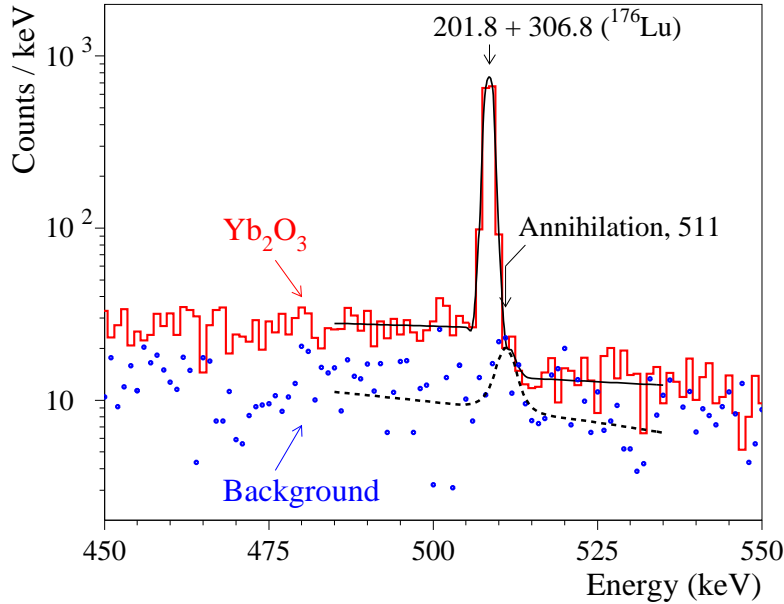,height=8.0cm}}
\caption{Parts of the energy spectra measured with the ytterbium
oxide sample over 2074 h (histogram) and background over 1046 h
(points) in the vicinity of the annihilation peak. The energies of
the $\gamma$ peaks are in keV.}
 \label{fig:511}
 \end{center}
 \end{figure}

We considered $\varepsilon\beta^{+}$ decay of $^{168}$Yb with the
transition to the ground and to the first 2$^+$ (79.8 keV) excited
states of $^{168}$Er, taking into account the maximal kinetic
energy available for the emitted positron (up to $\approx$387 keV,
depending on the binding energy of the atomic shell of the
daughter erbium atom). Irrespective to the $2\nu$ and $0\nu$ modes
and/or the transition level, the highest detection efficiency in
these processes is expected for a 511 keV $\gamma$ peak, populated
by two 511 keV $\gamma$ quanta emitted in the positron-electron
annihilation. The fits of the background spectra of the HPGe
detector acquired with the Yb$_2$O$_3$ powder and without sample
(see Fig. \ref{fig:511}) find the area of the annihilation peak
($25\pm13$) and ($19\pm18$) counts, respectively. We should add to
the model a sum peak with energy $(201.8+306.8)$ keV$=508.6$ keV
caused by the $^{176}$Lu internal contamination of the Yb$_2$O$_3$
sample. Taking into account the time of measurements of the sample
and of the background data, ($-12\pm38$) events can be ascribed to
the extra rate of the 511 keV peak in the data accumulated with
the ytterbium oxide sample, resulting in the $\lim S$ value equal
to 51 counts. Taking into account the rather similar detection
efficiencies for 511 keV $\gamma$ quanta emitted in the
$\varepsilon\beta^{+}$ decay processes in $^{168}$Yb, we obtained
for the both $2\nu$ and $0\nu$ modes of $\varepsilon\beta^+$ decay
the same half-life limit $\lim T_{1/2}= 2.8\times10^{17}$ yr.

\nopagebreak

\subsection{$2\beta^-$ decay of $^{176}$Yb to the first excited level of $^{176}$Hf}

The double beta decay of $^{176}$Yb is energetically allowed to
the ground and to the first $2^+$ excited (88.3 keV) states of
$^{176}$Hf, but only the transition to the excited level could be
investigated in the present experiment. The energy spectrum
measured with the Yb$_2$O$_3$ sample was fitted in the energy
interval $76-96$ keV by a model built of two Gaussian functions to
describe the effect searched for and the background 84.4 keV
$\gamma$ peak of $^{228}$Th, and a first degree polynomial as a
background model. The fit of the data (see Fig. \ref{fig:2n2K})
finds an area of the 88.3 keV peak of $239\pm26$ counts. However,
taking into account the contamination of the sample by $^{176}$Lu,
$181\pm13$ counts should be ascribed to the $^{176}$Lu activity.
The difference is $58\pm 29$ counts; although  this difference is
2 sigma from zero, it cannot be interpreted as the effect searched
for. Thus, the number of events excluded at 90\% C.L. according to
\cite{Feldman:1998} is $\lim S=106$ counts. Taking into account
the number of $^{176}$Yb nuclei in the  sample
($1.473\times10^{23}$) and the detection efficiency
$\eta^{2\nu}=1.958\times10^{-4}$
($\eta^{0\nu}=1.907\times10^{-4}$), we set the following half-life
limits for the $2\beta^-$ decay of $^{176}$Yb to the first $2^+$
excited state of $^{176}$Hf: $T_{1/2}^{2\nu(0\nu)}\geq
4.5(4.3)\times10^{16}$ yr. The achieved limit is weaker than the
one ($T_{1/2}^{2\beta^-}\geq 1.6\times 10^{17}$ yr at 68\% C.L.)
obtained in the experiment \cite{Derbin:1996} with a similar
technique for the sum of the $2\nu$ and $0\nu$ modes of the decay.
The reason for the worse sensitivity of the present experiment is
the contamination of the sample by lutetium.

\section{CONCLUSIONS}

The double beta decay processes in $^{168}$Yb have been
investigated for the first time using a highly purified 371 g
Yb$_2$O$_3$ sample and a ultra-low-background 465 cm$^3$ HPGe
$\gamma$ spectrometer. The experiment has been carried out at the
STELLA facility of the Gran Sasso underground laboratory over 2074
h.

The limits on the double beta decay of $^{168}$Yb to the ground
and excited states of $^{168}$Er were set at the level of $\lim
T_{1/2}\sim10^{14}$--10$^{18}$ yr. A possible resonant
neutrinoless double-electron capture in $^{168}$Yb to the
$(2)^{-}$ 1403.7 keV excited level of $^{168}$Er was restricted to
$\lim T_{1/2}=1.9\times 10^{18}$ yr. The utilization of ytterbium
with natural isotopic composition, containing only $\sim$0.1\% of
$^{168}$Yb, and the passive source technique (only few \%
detection efficiency) considerably limited the sensitivity. In
particular, the achieved sensitivity is a few orders of magnitude
worse in comparison to $\lim T_{1/2}\sim$ 10$^{21}$--10$^{22}$ yr
achieved in the most sensitive experiments looking for double beta
processes with charge increase, namely double-electron capture,
electron capture with positron emission, and double-positron decay
(references for these experiments can be found in
\cite{Belli:2018}). The half-life of the $2\beta^-$ decay of
$^{176}$Yb to the $2^+$ 88.3 keV excited level of $^{176}$Hf was
restricted to $T_{1/2}^{2\nu(0\nu)}\geq4.5(4.3)\times10^{16}$ yr.
The sensitivity is slightly worse than that in the previous
experiment \cite{Derbin:1996} due to the strong contamination of
the Yb$_2$O$_3$ sample with $^{176}$Lu on the level of $\sim0.4$
Bq/kg.

The present sensitivity to double beta decay of $^{168}$Yb and
$^{176}$Yb can be significantly improved using isotopically
enriched material and an active source technique. An Yb-loaded
liquid scintillator (see e.g. \cite{Raghavan:1997}), or an
Yb-containing crystal scintillator (e.g. Yb-doped YAG
\cite{Antonini:2002}), or a new scintillation material with Yb
could be viable possibilities for the realization of a
source-equal-detector double beta decay experiment.

A liquid-liquid extraction based method to reduce the
contamination of the Yb$_2$O$_3$ sample in potassium, radium and
thorium by 3, 34 and 68 times, respectively, has been developed.
These results could pave the way to the use of ytterbium in a
large-scale experiment aiming at the investigation of neutrino
oscillations and/or double beta decay
\cite{Raghavan:1997,Zuber:2000}. A rather high contamination of
the Yb$_2$O$_3$ sample by lutetium could be reduced by selection
of materials with different mineral origin. For instance, we have
observed an order of magnitude variation of the $^{176}$Lu
activity ($\sim2$ mBq/kg and $\sim20$ mBq/kg) in two samples of
gadolinium oxide from different suppliers. This is despite the
fact that lower lutetium contamination was observed in the
material of a lower purity grade. Furthermore, the samples of
cerium, erbium and neodymium (elements chemically similar to
ytterbium) utilized in the double beta decay experiments
\cite{Belli:2017,Belli:2018,Barabash:2018} are characterized by a
much less $^{176}$Lu activity on the level of $0.3-4$ mBq/kg.
Therefore, one should organize screening of materials of different
origin before final purification of samples for a low counting
experiment. Further purification of ytterbium from lutetium could
be achieved by using chromatographic separation
\cite{Kumar:1994,Max-Hansen:2011,Ojala:2012}. Development of the
chromatographic purification method is in progress.

\section{ACKNOWLEDGEMENTS}

The group from the Institute for Nuclear Research (Kyiv, Ukraine)
was supported in part by the program of the National Academy of
Sciences of Ukraine ``Fundamental research on high-energy physics
and nuclear physics (international cooperation)''. O.G.P. was
supported in part by the project ``Investigations of rare nuclear
processes'' of the program of the National Academy of Sciences of
Ukraine ``Laboratory of young scientists''.


\begin{thebibliography}{99}
 \bibitem{Tretyak:2002} V.I.~Tretyak, Yu.G.~Zdesenko, Tables of double $\beta$ decay data -- an update, At. Data Nucl. Data Tables 80 (2002) 83.
 \bibitem{Barabash:2015} A.S.~Barabash, Average and recommended half-life values for two-neutrino double beta decay, Nucl. Phys. A 52 (2015) 935.
 \bibitem{Tanabashi:2018} M.~Tanabashi et al. (Particle Data Group), Phys. Rev. D 98 (2018) 030001.
 \bibitem{Vergados:2016} J.D. Vergados, H. Ejiri, F. Simkovic, Neutrinoless double beta decay and neutrino mass, Int. J. Mod. Phys. E 25 (2016) 1630007.
 \bibitem{Delloro:2016} S.~Dell'Oro, S.~Marcocci, M.~Viel, F.~Vissani, Neutrinoless Double Beta Decay: 2015 Review, AHEP 2016 (2016) 2162659.
 \bibitem{Bilenky:2015} S.M.~Bilenky, C.~Giunti, Neutrinoless double-$\beta$ decay: A probe of physics beyond the standard model, Int. J. Mod. Phys. A 30 (2015) 1530001.
 \bibitem{Ratkevich:2017} S.S.~Ratkevich et al., Comparative study of the double-$K$-shell-vacancy production in single- and double-electron-capture decay, Phys. Rev. C 96 (2017) 065502.
 \bibitem{Belli:2018} P.~Belli et al., First search for $2\varepsilon$ and $\varepsilon\beta^+$ decay of $^{162}$Er and new limit on $2\beta^{-}$ decay of $^{170}$Er to the first excited level of $^{170}$Yb, J. Phys. G: Nucl. Part. Phys. 45 (2018) 095101.
 \bibitem{Kotila:2012} J.~Kotila and F.~Iachello, Phase-space factors for double-$\beta$ decay, Phys. Rev. C 85 (2012) 034316.
 \bibitem{Kotila:2013} J.~Kotila and F.~Iachello, Phase-space factors for $\beta^+\beta^+$ decay and competing modes of double-$\beta$ decay, Phys. Rev. C 87 (2013) 024313.
 \bibitem{Hirsch:1994} M.~Hirsch, K.~Muto, T.~Oda, H.V.~Klapdor-Kleingrothaus, Nuclear structure calculation of $\beta^+\beta^+$, $\beta^+$/EC and EC/EC decay matrix elements, Z. Phys. A 347 (1994) 151.
 \bibitem{Deppisch:2012} F.F.~Deppisch, M.~Hirsch, H.P\"{a}s, Neutrinoless double-beta decay and physics beyond the standard model, J. Phys. G: Nucl. Part. Phys. 39 (2012) 124007.
 \bibitem{Pas:2015} H.P\"{a}s, W.~Rodejohann, Neutrinoless double beta decay, New J. Phys. 17 (2015) 115010.
 \bibitem{Winter:1955} R.~Winter, Double $K$ Capture and Single $K$ Capture with Positron Emission, Phys. Rev. 100 (1955) 142.
 \bibitem{Voloshin:1982} M.B.~Voloshin, G.V.~Mitselmakher, R.A.~Eramzhyan, Conversion of an atomic electron into a positron and double $\beta^+$ decay, JETP Lett. 35 (1982) 656.
 \bibitem{Bernabeu:1983} J.~Bernabeu, A.~De~Rujula, C.~Jarlskog, Neutrinoless double electron capture as a tool to measure the electron neutrino mass, Nucl. Phys. B 223 (1983) 15.
 \bibitem{Sujkowski:2004} Z.~Sujkowski, S.~Wycech, Neutrinoless double electron capture: A tool to search for Majorana neutrinos, Phys. Rev. C 70 (2004) 052501(R).
 \bibitem{Meija:2016} J.~Meija et al., Isotopic compositions of the elements 2013 (IUPAC Technical Report), Pure Appl. Chem. 88 (2016) 293.
 \bibitem{Eliseev:2011} S.~Eliseev et al., $Q$ values for neutrinoless double-electron capture in $^{96}$Ru, $^{162}$Er, and $^{168}$Yb, Phys. Rev. C 83 (2011) 038501.
 \bibitem{Wang:2017} M.~Wang et al., The AME2016 atomic mass evaluation, (II). Tables, graphs and references, Chin. Phys. C 41 (2017) 030003.
 \bibitem{Baglin:2010} C.M.~Baglin, Nuclear Data Sheets for A = 168, Nuclear Data Sheets 111 (2010) 1807.
 \bibitem{Basunia:2006} M.S.~Basunia, Nuclear Data Sheets for A = 176, Nuclear Data Sheets 107 (2006) 791.
 \bibitem{Ceron:1999} V.E. Ceron, J.G. Hirsch, Double electron capture in $^{156}$Dy, $^{162}$Er and $^{168}$Yb, Phys. Lett. B 471 (1999) 1.
 \bibitem{Krivoruchenko:2011} M.I.~Krivoruchenko, F.~\v{S}imkovic, D.~Frekers, A.~Faessler, Resonance enhancement of neutrinoless double electron capture, Nucl. Phys. A 859 (2011) 140.
 \bibitem{Derbin:1996} A.V.~Derbin, A.I.~Egorov, V.N.~Muratova, S.V.~Bakhlanov, New limits on half-lives of $^{154}$Sm, $^{160}$Gd, $^{170}$Er, and $^{176}$Yb with respect to double $\beta$ decay to the excited $2^+$ states of daughter nuclei, Phys. At. Nucl. 59 (1996) 2037.
 \bibitem{Raghavan:1997} R.S.~Raghavan, New Prospects for Real-Time Spectroscopy of Low Energy Electron Neutrinos from the Sun, Phys. Rev. Lett. 78 (1997) 3618.
 \bibitem{Zuber:2000} K.~Zuber, Double beta decay with large scale Yb-loaded scintillators, Phys. Lett. B 485 (2000) 23.
 \bibitem{Park:2008} H.~Park et al., Stable Isotope Production of $^{168}$Yb and $^{176}$Yb for Industrial and Medical Applications, J. Nucl. Sci. Tech. 45, suppl. 6 (2008) 111.

 \bibitem{Laubenstein:2017} M.~Laubenstein, Screening of materials with high purity germanium detectors at the Laboratori Nazionali del Gran Sasso, Int. J. Mod. Phys. A 32 (2017) 1743002.
 \bibitem{Agostinelli:2003} S.~Agostinelli et al., GEANT4---a simulation toolkit, Nucl. Instrum. Meth. A 506 (2003) 250.
 \bibitem{Allison:2006}  M. Boswell et al., IEEE-NS 58(3) (2011) 1212; J.~Allison et al., Geant4
developments and applications, IEEE Trans. Nucl. Sci. 53 (2006) 270.
 \bibitem{Ponkratenko:2000} O.A.~Ponkratenko et al., Event generator DECAY4 for simulating double-beta processes and decays of radioactive nuclei, Phys. At. Nucl. 63 (2000) 1282.
 \bibitem{DECAY0} V.I.~Tretyak, private communication.

 \bibitem{Boiko:2017} R.S.~Boiko, Chemical purification of lanthanides for low-background experiments, Int. J. Mod. Phys. A 32 (2017) 1743005.
 \bibitem{Belli:2014} P.~Belli et al., Search for double beta decay of $^{136}$Ce and $^{138}$Ce with HPGe gamma detector,
Nucl. Phys. A 930 (2014) 195.

 \bibitem{Kawrakow:2003} I.~Kawrakow, D.W.O.~Rogers, The EGSnrc code system: Monte Carlo simulation of electron and photon transport, NRCC Report PIRS-701, Ottawa, 2003.
 \bibitem{Feldman:1998} G.J.~Feldman, R.D.~Cousins, Unified approach to the classical statistical analysis of small signals, Phys. Rev. D 57 (1998) 3873.

 \bibitem{Firestone:1998} R.B.~Firestone et al., \textit{Table of Isotopes}, 8-th ed., John Wiley, New York, 1996 and CD update, 1998.
 \bibitem{Belli:2012} P.~Belli et al., Search for double-$\beta$ decay processes in $^{106}$Cd with the help of a $^{106}$CdWO$_4$ crystal scintillator, Phys. Rev. C 85 (2012) 044610.

 \bibitem{Antonini:2002} P.~Antonini et al., Properties of Yb:YAG scintillators, Nucl. Instrum. Meth. A 486 (2002) 220.
 \bibitem{Belli:2017} P.~Belli et al., New limits on 2$\varepsilon$, $\varepsilon\beta^+$ and 2$\beta^+$ decay of $^{136}$Ce and $^{138}$Ce with deeply purified cerium sample, Eur. Phys. J. A 53 (2017) 172.
 \bibitem{Barabash:2018} A.S.~Barabash et al., Double beta decay of $^{150}$Nd to the first excited 0$^+$ level of $^{150}$Sm: preliminary results, Nucl. Phys. At. Energy 19 (2018) 95.
 \bibitem{Kumar:1994} M.~Kumar, Recent trends in chromatographic procedures for separation and determination of rare-earth elements, A Review, Analyst 119 (1994) 2013.
 \bibitem{Max-Hansen:2011} M.~Max-Hansen et al., Optimization of preparative chromatographic separation of multiple rare earth elements, J. Chromatogr. A 1218 (2011) 9155.
 \bibitem{Ojala:2012} F.~Ojala et al., Modelling and optimization of preparative chromatographic purification of europium. J. Chromatogr. A 1220 (2012) 21.

\end{thebibliography}
\end{document}